# From Forced Working-From-Home to Working-From-Anywhere: Two Revolutions in Telework

Darja Šmite[†]
Blekinge Institute of Technology
Karlskrona, Sweden
darja.smite@bth.se

Nils Brede Moe
SINTEF
Trondheim, Norway
nils.b.moe@sintef.no

Eriks Klotins
Blekinge Institute of Technology
Karlskrona, Sweden
eriks.klotins@bth.se

Javier Gonzalez-Huerta
Blekinge Institute of Technology
Karlskrona, Sweden
javier.gonzalez.huerta@bth.se

## ABSTRACT

The COVID-19 outbreak has admittedly caused a major disruption worldwide. The interruptions to production, transportation, and mobility have clearly had a significant impact on the well-functioning of the global supply and demand chain. But what happened to the companies developing digital services, such as software? Were they interrupted as much or at all? And how has the enforced Working-From-Home (WFH) mode impacted their ability to continue to deliver software? We hear that some managers are concerned that their engineers are not working effectively from home, or even lack the motivation to work in general, that teams lose touch and that managers do not notice when things go wrong. In this article, we share our findings from monitoring the situation in an international software company with engineers located in Sweden, USA, and the UK. We analyzed different aspects of productivity, such as developer satisfaction and well-being, activity, communication and collaboration, efficiency and flow based on the archives of commit data, calendar invites, and Slack communication, as well as the internal reports of WFH experiences and 18 interviews. We find that company engineers continue committing code and carry out their daily duties without significant disruptions, while their routines have gradually adjusted to "the new norm" with new emerging practices and various changes to the old ones. In a way, our message is that there is no news, which is good news. Yet, the experiences gained with the WFH of such scale have already made significant changes in the software industry's future, work from anywhere (WFX) being an example of major importance.

## 1 The Work-From-Home Mode

One of the first studies of pioneers working from home (WFH) or doing home telework published in 1984 [1] starts with a futuristic vision of the white-collar labor force working in home offices. Just 36 years later, IT industry players like Facebook and Microsoft make revolutionary announcements of new remote work policies and open remote positions, while others like Twitter, Salesforce and Spotify make further steps introducing permanent work-from-anywhere policies based on the forced WFH experiences during the pandemic. What is the role of "working-from-home," and why the experiences during the fight against the COVID-19 are likely to bring tremendous changes into the workplace?

Working from home is not a new phenomenon. The first studies originate in the 70s along with the declining costs of data communications and the influx of microcomputers into homes and offices [1, 2]. Back then, motivation to telework was either related to the energy shortages and the willingness to decrease the daily commute or associated with a clear gender segregated motivation. The typical teleworkers at the time included *"self-disciplined full-time clerical women seeking income at reduced personal expense, managerial and professional mothers wanting to nurture young children without dropping completely behind in their careers; and male managers or professionals who value the part-time integration of work and family life more than they do competition for further advancement in their organizations"* [1]. Since then, many things have changed, like the accessibility and advancement of telecommunication services and equipment, and the need for employee empowerment and improved work-life balance.

Quite a few companies today implement WFH as an element of flexibility leading to a regular but partial practice (a few days a week) [3] and some even as a general company practice [4]. Telework is often associated with the perceived increase of productivity and job satisfaction, mostly self-reported by home workers, and a great managerial issue [2]. Managers repeatedly raise the question of whether "working-from-home" would not lead to "shirking from home" [5]. The typical "Theory X style managers" [6] with a low perception of self-efficacy, i.e., who do not rely on their employees' ability to handle remote infrastructure, solve situations independently, manage time properly or working without supervision, have a skeptical attitude towards telework [7]. The mistrust is also felt by the teleworkers who confirm that their supervisors tolerate their preference for WFH unwillingly [1]. Some coworkers share skeptical attitudes towards teleworkers, thinking that the off-premises colleagues are not working full time, while others show acceptance or express awe, envy, jealousy, or resentment [1].

Motivated by the willingness to understand the actual state and future of telework, we share our findings from studying **the changes in individual productivity and work routines in the first eleven weeks of WFH** after our case company closed their offices.

## 2   Case Company: InterSoft

InterSoft (pseudonym used for anonymity) is an international software company with hundreds of teams working in development offices in Sweden, the UK and USA delivering millions of lines of code of complex software per year. InterSoft is a modern agile company with advanced ways of working that promote collaboration and teamwork. InterSoft cultivates the culture of self-management and increased autonomy with decentralized decision-making structures. Thanks to these aspects and the geographic distribution, InterSoft has had the facilitating conditions and infrastructure to enable distributed work already before the pandemic (usually associated with both the ease to implement telework, and the positive attitude toward it [7]). In March 2020, all InterSoft employees in all locations were instructed to work from home, prohibiting access to the office spaces, initially for two weeks, which during our study was extended four times and by the time of writing reached September 2021. Recently, InterSoft has announced a new policy to enable *Working from Anywhere* (WFX). Notably, nobody could foresee that the WFH mode would last so long. To support employees in transition to WFH, InterSoft launched a program for reimbursing home office equipment and acquired various remote collaboration software licenses in the early weeks of WFH and organized numerous experience-sharing activities and fora. Company management exhibited high levels of trust in the individuals and teams' ability to cope with uncertainty and adjust to the situation [6].

All this support made InterSoft an excellent case for our study since many companies can learn from them. Besides, InterSoft was involved in another research study, and thus we were already familiar with the context and had access to various data sources important for our research.

## 2   Overview of the Study

Our goal is to understand how engineers cope with the WFH mode and what changes in daily routines and productivity have happened. The enforced WFH mode in our study is unique compared to prior studies of telework, which are biased towards voluntary teleworkers [2]. We also address a common definitional issue of who qualifies as a teleworker, since many previous studies focused on individuals following the practice only partially [2]. Our study is driven by the following research question: How is individual productivity affected by WFH? To answer our question, we used the mixed method employing concurrent procedures [23] by converging quantitative and qualitative data to provide a comprehensive analysis of the research problem. In our study, we combined data from a variety of sources, including GIT commits, Slack posts, calendar invitations, and 18 semi-structured interviews with 15 engineers and three managers. Interviewees were selected by convenience sampling to have representatives from the main locations, age groups, seniors and juniors in the company, and with different family situations (living alone, with a spouse, with kids), as also studied in prior research on teleworkers [1]. All interviews were 45-60 min long, conducted by two researchers in English via Zoom. All interviews but one were audio-recorded with the consent of the interviewees. One of the interviewers led the interview, while the other took detailed notes (close to transcription). After the interview, notes were refined and complemented if necessary. Finally, we had access to four internal company reports of the WFH experiences, which served as 1) input for identifying interesting questions and categories for our analysis, and 2) an additional source of company-wide inquiry, verifying some of our findings by representing a wider sample.

To answer our question regarding pandemic productivity of software engineers working from home, we combined the findings from a thematic analysis of the qualitative data with the quantitative data analysis. **Productivity** in our study is defined as a complex multifaceted concept described across five dimensions [18]: satisfaction and well-being, performance, activity, communication and collaboration, and efficiency and flow, which served as higher level categories in thematic analysis. Instead of using any single productivity measure, which all have been criticized [18], we decided to rely on a combination of quantitative and qualitative data that explains the changes in each of these dimensions comparing WFH to the work in the office. In the following, we explain our data collection and analysis in more detail.

To assess changes in **satisfaction and well-being**, we asked engineers and managers to describe their overall attitude and *satisfaction* with their workplace at home, relationship with the team members, and tool support for various daily activities. We also looked at the work in unusual work hours through different activities (code commits, meetings and Slack communication) during the day. Further, we asked whether the interviewees would prefer to work from home in the future, or rather return to the office, as another indicator of satisfaction. We also asked the interviewees to drive us through their daily routines and checked for the signs of *well-being* such as healthy lifestyle, overall happiness or, on the contrary, emotional problems [19].

To assess changes in **activity**, we looked at the count of outputs completed in the course of performing work [18] by analyzing when and how much code engineers commit to the main branch in all version control repositories. We compared the relative distribution of commits during the day during 2020-01-01 – 2020-03-10 as the "office" control period and during 2020-03-11 – 2020-05-30 as the WFH period. The changes in commit volume have also been compared to 2019. Data cleaning included filtering out the top 25% of the commits considering size, because these are likely to be committed by automated *bots*.

To assess changes in **communication and collaboration**, i.e. the way individuals and teams communicate and work together [18], we asked the interviewees to describe how their teams adjusted to the remote work and also analyzed quantitative data about the changes in electronic communication using the number of Slack posts in public channels with >10 members comparing 2020-01-01 – 2020-03-10 as the "office" period and 2020-03-11 – 2020-05-30 as the WFH period.

To measure changes in **efficiency and flow**, we asked the interviewees to comment on their perceived ability to stay in flow and quantitatively analyzed interruptions introduced by meetings. Quantitative data comprised meeting invitations accepted by engineers, non-recurring and recurring. Data cleaning included filtering out: meetings longer than 8 hours as events that do not reflect the meeting habits; and meetings with less than 2 participants or more than 20 as self-bookings and large gatherings. We compared the daily routines and analyzed the changes in the average number of meetings, the average duration of meetings, the total time spent in meetings per engineer per week, and the total number of meetings scheduled per week during the "office" control period and during the WFH period.

Notably, in this paper, we do not report any changes in **Performance**. This is because we did not have access to any reliable data sources that would let us assess work *outcomes* rather than *outputs* (which we linked to engineers' activity) such as code quality or customer satisfaction as suggested in [18].

We based our analysis on obfuscated datasets cleaned from the information that would allow identifying a person. The quantitative analysis was performed from the point of view of the user performing different actions (commits, meetings or Slack messages) happening in the user's timezone. Therefore, the same events, e.g. a meeting, are accounted as occurring in different hours for users in different time zones, depending on their location, although happening simultaneously. The quantitative analysis of these datasets was performed using NumPy, Scipy, Pandas and MatPlotLib libraries for Python 3.7.4.

## 3  Productivity in the Forced WFH Mode

In the following, we capture the WFH experiences during the first eleven weeks (w11-22) based on interviews and the quantitative data collected from the company repositories.

### 3.1    Satisfaction and well-being

**Ergonomics**: From the start, many employees realized that their homes do not provide sufficiently ergonomic work conditions. They used the company support to buy an external monitor and keyboard, an office chair, an external keyboard, or other missing equipment. Many of the everyday routines had to adjust too.

Transition to distributed work: The analysis of the daily activities at InterSoft based on GIT commits, calendar invitations, and Slack communication in public channels shows that engineers follow similar daily routines compared to working in the office (see Figure 1). One reason for what can be regarded as a smooth transition into the WFH is related to the fact that InterSoft has operated in multiple locations for many years and is experienced with distributed collaborative work and virtual teams. Another reason is the high level of team authority at InterSoft, so operational decisions are not hindered by such outstanding events as the COVID-19 outbreak.

**Work hours**: While many routines stay similar as in the office times, we observe that engineers start the day a bit earlier and end the day a bit later, about 30-60 min on average (see Figure 1). As one engineer explained: *"I usually get up, take a glass of water, and just start working. It's what differs now – I don't take a shower or eat breakfast"*. When discussing the daily routines and comparing the quantitative data analysis week by week, we noticed that some engineers worked longer days, as also found in an extensive study of GitHub projects [8] and an early WFH study published by Bloomberg based on the first week's user activity data from the network service provider NordVPN Teams [9]. Some engineers we talked to said they have been unable to distinguish work and personal life and continued working late or exchanged emails and Slack messages in the evenings. A manager confirmed: *"I see a lot of people tired"*. One practice we found to address this was a "hard stop", i.e., a preset time for logging off or turning off the computer. Many people who had a problem stopping working late said to have shifted to a stricter routine after the first 2-4 weeks of the WFH.

**Physical activity during the day:** The exhaustion comes not only from just working longer hours. Our data shows that the lunch breaks became shorter (increased activity between 12-13 o'clock in Figure 1). Many engineers we spoke to confessed that they are not good at taking breaks. The usual breaks and positive interruptions from colleagues at the office or invitations to grab a coffee or a snack in the kitchen were suddenly gone. Even attempts to have a joint virtual break are not always successful, as an engineer explains: *"If you write someone on email, "Hey, are you free?" There is always a delay, then you wait for 10 minutes, and then you are back to work again. It would be nice to take more breaks, but I want to take breaks and talk to people"*. Staying at home, especially in solitude, for many means that the walk to the kitchen or the restroom and back takes just a few minutes, after which you are back at the work desk. Thus, the first weeks of the WFH showed a significant decrease in physical activity without extra efforts (not to mention the lack of gym exercise for many out of precaution or because of the lockdown).

**Loneliness:** Emotional and social isolation often cause loneliness. Many of our interviewees admitted that their social life due to the lockdown or social distancing measures suffered. Loneliness was especially tough for single expats, who lived abroad from their families. However, not everyone experiences reduced social interaction as a problem. Some people, especially those recovering from a recent burnout, enjoyed working in isolation, which was associated with having more control over their workday. Similarly, prior research suggests that the lack of social contacts will not necessarily make everybody feel lonely, but only those having a high desire for social relations [10].

### 3.2    Activity

Code production: To understand how WFH has affected engineers' performance, we analyzed the total lines of code committed to version control main branches per week – total for the company and normalized per developer (see Figure 2). The 2020 data covering the transition to the WFH mode (the red line) shows a dip in code production in the first three weeks of WFH. Comparing the code production figures in the first 10 weeks of 2020 with the WFH period, normalized per developer shows that, on average, engineers merge slightly more code (+6%) in the WFH in comparison with the work in the office (see the red dot in Figure 2). In general, it is fair to note that the shape of lines is very similar, which means that WFH did not have any pronounced effect on the output measured in terms of lines of code produced merged to the main branch.

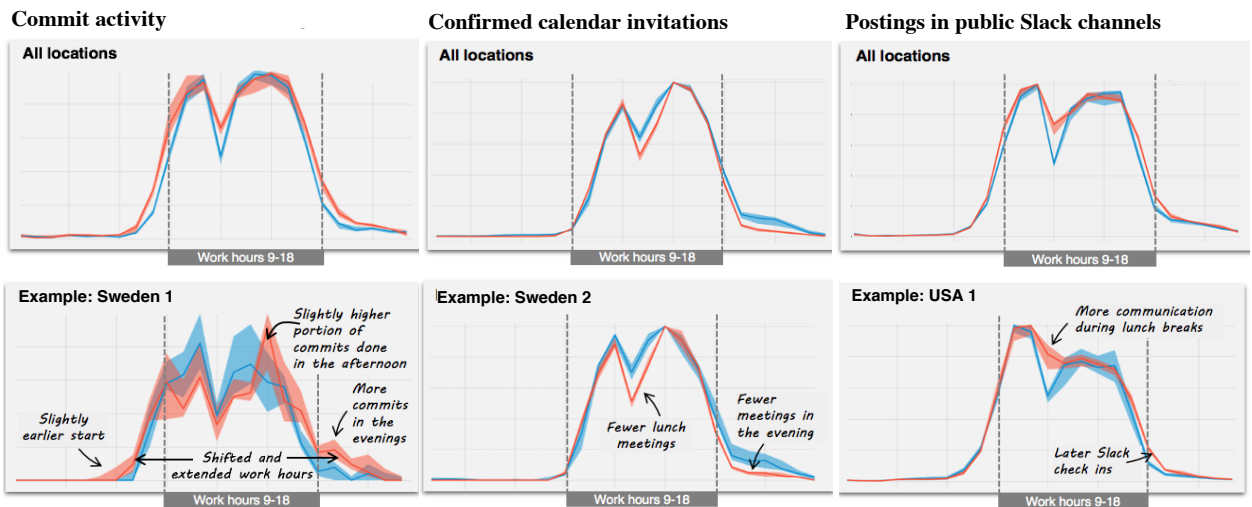

**Figure 1: Daily activity.** The blue area represents 25-75% quantiles during the office work (2020-01-01 – 2020-03-11) with the blue line denotes the average value, while the red area and line represent the WFH period (2020-03-12 – 2020-05-30)

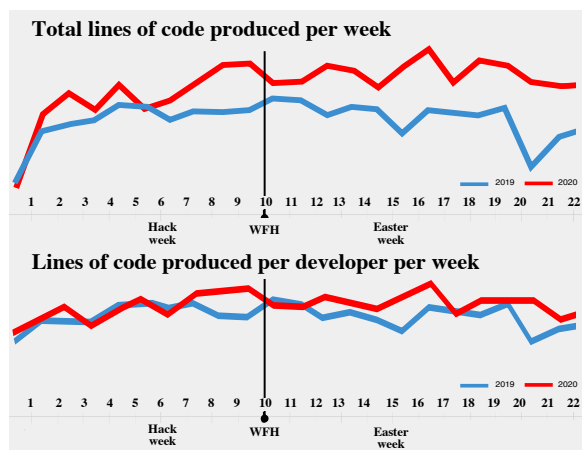

**Figure 2: Total LOC produced per week.**

### 3.3 Communication and collaboration

**Socialization and informal communication**: Since WFH meant that people could no longer meet physically in the office, we learned that those in need of social contact sought ways to interact virtually. Conversations by the coffee machines and kitchen gathering traditions are deep-rooted in the daily routines of InterSoft. Spontaneous interactions are used to discuss work, resolve problems, learn something new or just catch up on non-work-related topics. While working remotely, all teams and departments have tried to carry out virtual coffee breaks at least once a week, although with mixed success. Many complained that the ease of socialization is gone, and there was often an awkward silence. Some joked that with the lack of sweet pastry, the extra motivation to attend coffee breaks disappeared. As a result, the virtual hangouts often had few participants. Therefore, it's not surprising that almost all interviewees independently reported that their daily stand-ups were prolonged to facilitate informal communication since these meetings became the main or even the only contact times, at least for some team members.

**Team cohesion**: Many engineers discussed that the level of socialization in the teams dropped. Some revealed they had a feeling of becoming less connected with their teammates, while others, on the contrary, said they feel more connected than before. The latter was because engineers could see the glances of the family life during the video sessions, which they have not seen before. Challenges with keeping the team spirit have become the regular topic during team retrospectives. Many teams sought and quickly adjusted new approaches, such as regular video-game sessions, scheduled drop-in video channels for team members, regular short virtual *"Hey, good morning"* greetings, shared lunches, fruit breaks and afterwork in front of the video chat.

**Problem-solving:** We learned that spontaneous discussions are important not only for socializing, but also for problem-solving. Our interviewees all echoed each other saying that there are no "over-the-shoulder" conversations anymore, for good and for bad. On the one hand, this meant that one could no longer shout out a question and get a quick answer. Thus, engineers can be blocked for a longer time than usual. On the other hand, many felt less distracted by peers and benefitted from the uninterrupted flow (we discuss this later). Solving problems individually, especially for novices, led to becoming more independent. In the end, some speculated that the effect of having a better flow and being stuck for longer might compensate each other productivity wise.

**Knowledge sharing**: Some of the "over-the-shoulder" conversations have moved to Slack chats. As a recent hire explains: *"I usually gather perspectives and ideas from others, and then start working on the problem. When I get stuck, I ask others about possible solutions. Now it is a little bit different. If … I am unsure* *how to do (a task), I usually ping someone in Slack who has done something similar and have a brief discussion. If I need more ideas, I can ping the same person or post it on the team Slack channel. You have to know who to ping".* The Slack records show a significant increase in the number of slack messages posted per week in the public channels alone (see Figure 3).

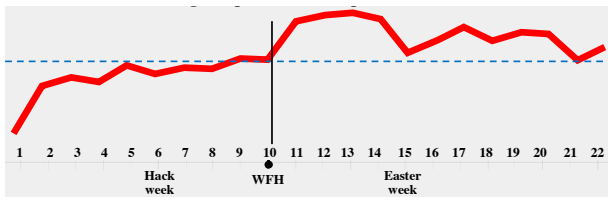

**Figure 3: Number of Slack messages in public channels.**

Pairing: The practices of pairing and mob programming suffered similarly to all interactions that have been previously held spontaneously. Even big fans of pairing reveal that they hardly pair remotely. As an engineer explains: *"(WFH) is a barrier to doing pair programming or mob programming. It's easier in the office. We've tried it once since we started working from home, and I think it worked quite well, but I think it's hard. I am not sure why. I've tried to say that I want to pair, and then it's 'Yeah, sure. Let's do it after lunch'. And then things change after lunch, and then you don't do it. Maybe it's because we need to schedule this more explicitly".* One solution has been to pair while doing parallel work with frequent synchronization. As another engineer explains: *"Now we sync several times a day, without sharing the screen. When needed, we say 'Are you free?' and either jump on a call or start screensharing".* Many have successfully used Zoom or Mural (a tool for remote collaboration). Interestingly, one engineer reported doing more pairing in their team during WFH. These were daily dedicated sessions (between the stand-up and the lunch, sometimes continuing after lunch) with an open video channel. Find more about WFH pair programming at InterSoft in [20].

**Communication in partially distributed teams:** Many members of teams that have been split across InterSoft locations said to experience better times than before the WFH. As an engineer explained: *"Everyone is now on the same terms. [...] Now I collaborate as much with [Location 1] people as with [Location 2] people".* Meetings in split teams and department meetings are said to happen on an equal basis with everybody dialing in. There are no obvious cliques in a meeting room and no location-specific disadvantages since the physical group division and co-location do not exist anymore. Some engineers even said that they perceive everybody (within the same time zone) to be equally accessible, removing the distance between offices and the floors of the same office, and report increased communication with people they were unfamiliar with before.

### 3.4 Efficiency and flow

**Perceived ability to stay in flow**: Since the spontaneous interruptions from colleagues and some of the other reasons for taking regular breaks during WFH disappeared, many engineers reported feeling an increased level of focus and "being in the flow". As an engineer explains: *"There is a great sense of productivity when you get into that momentum".* These were primarily engineers who worked on independent tasks and were familiar with the work they did. This can be also indirectly confirmed by comparison of the output measures of the amount of code individual engineers produced before and after WFH. When we calculated the number of developers producing more code than before WFH versus those producing less, we came up with the 60:40 proportion. The downside of having the increased focus was that engineers felt exhausted earlier than normally. But the feeling of gained productivity still appears to be more important, as explained by the same engineer: *"It is not necessarily a bad thing. A more focused but shorter workday is nice".*

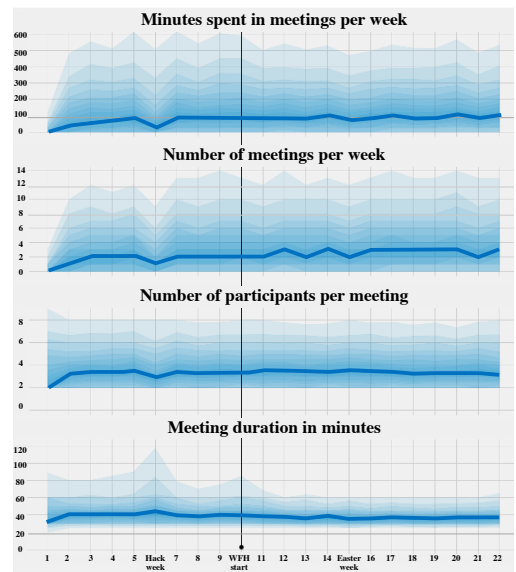

**Figure 5:** Changes in meeting patterns for engineers. Shades of blue denote 5%-95% quantiles. The solid line is the median.

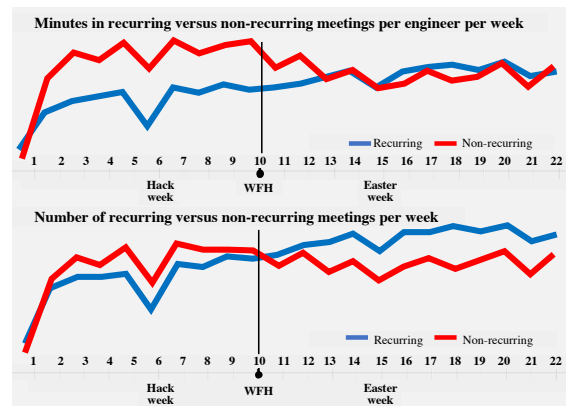

**Figure 6: Recurring versus non-recurring meetings.**

**Meetings**: Efficiency and flow require extended periods of concentration, which meetings can interrupt. A high number of meetings was found to cause stress during the pandemic [19]. We, therefore, decided to analyze calendar invitations for meetings no longer than 8h and involving 2-20 people. Figure 5 illustrates the changes in meeting patterns. When comparing the beginning of the year with the WFH period, we can see that median values across all plots have only negligible changes, if any. Yet, the outlining cases have some differences. For instance, the number of meetings per week slightly increased after transitioning to the WFH, perhaps due to a shift from spontaneous physical to scheduled online meetings. We also see a trend of meetings becoming shorter (no meetings longer than 1h) and engineers having fewer long meetings (changes for those who spent more than 5h in meetings per week). Engineers explained that at the beginning of WFH, everybody tried to mimic the office routines and carry on with usual meetings, soon realizing that online meetings are much more exhausting, as discussed in numerous WFH-related posts, e.g., on Zoom fatigue [21]. This was especially noticeable for those having several meetings in a row – since people are no longer changing rooms and switching between meetings requires pressing the "leave call" and then the "join call" buttons. Exhaustion is a possible explanation for why the meetings' duration became slightly shorter already in the second week of WFH. Major changes occur in the time spent in non-recurring meetings and the shift to recurring meetings (see Figure 6). Spontaneous conversations that would happen organically in the office space, such as virtual coffee breaks, socialization hours, or pair programming slots, in WFH became scheduled.

## 4 Conclusions and Lessons Learned

COVID-19 pandemic provided humanity with a unique opportunity to learn from experiencing the world with a forced working-from-home practice and the lack of common global travel and even local face-to-face interaction. The situation is unique because there is hardly any existing guidance since nobody knows how to manage the global lockdown of such a scale. In this paper, we documented experiences from one company and demonstrated the transition its employees made to adjust to the WFH (see our summary in Table 1). Our results show that software companies can indeed work remotely with no significant impact on various aspects of productivity. But what can we learn from these experiences?

### 4.1   What Did We Learn About Telework?

We analyzed different aspects of productivity, including developer satisfaction and well-being, activity, communication and collaboration, efficiency and flow. Generally, our results are consonant with related studies on WFH under COVID-19 that found pandemic productivity in software companies not much changed [19], and prior research suggesting that telework leads to increased productivity and satisfaction [2]. Yet, a critical finding in our study is that WFH is not for everybody, something that is evident also in the surveys of perceived productivity in WFH, which suggest that some are doing better while others are doing worse [22]. Like the study of WFH pioneers [1], we found that singles (especially expats) whose social life happens in the office do not favor WFH. Further, while some experienced challenges when balancing work and family life or focusing on the job when being at home (the group of not self-disciplined in [1]), we found that inability to avoid WFH forced many to find ways to overcome these challenges, adjusting their family routines or equipping a more isolated workspace. Thus, WFH appeared to be a way of working for many more people than initially thought. Additionally, we found new factors with perhaps more substantial impact on the individual motivation for WFH, such as supporting recovery from a burnout and a vested interest in remote work stemming from the eventual plans to move abroad or positive experiences from creating closer bonds with the family during the forced WFH.

### 4.2   Transition to WFH

We found the current form of WFH challenging, questioning WFH as a part of the future permanent practice. While activity, efficiency and flow at the moment of our study have not changed much (the changes have not surfaced yet), other productivity aspects such as well-being, communication and collaboration have been impacted. The list of practices and habitual routines that have disappeared in the WFH reality is long, including spontaneous interactions, ad-hoc meetings, peer support, whiteboards, "water cooler" conversations, shared coffee machines, cafeterias, lounges, gaming areas, along with serendipitous moments to exchange socio-emotional information, and opportunities to be introduced to unfamiliar colleagues. Even to a large degree the so common pairing and mob programming have significantly decreased [20]. While there is a belief that engineers can learn to facilitate organic communication virtually, spontaneous interactions between members of fully virtual teams require tools that provide rich group awareness (up-to-the-minute understanding of the whereabouts of the members and their actions) [15]. In our study, despite the rich tool support provided by InterSoft, such awareness was not evidenced. In Table 1 we summarize the practices, ceremonies and routines that have been changed or newly introduced to facilitate WFH, reflecting the significant changes that COVID-19 brought to the daily rhythm, meeting habits, teamwork and social interactions. Even though to date we have not found any significant changes in the engineers' outputs, the absence of office routines and the shortcomings of the virtual ones raise concerns for the long-term impacts of the WFH on the choice of solutions, ability to solve complex tasks together, initiating new unfamiliar things, developing new ideas, the speed of planning, onboarding new employees, and sharing knowledge. While the organizers of the virtual equivalents for many events at InterSoft have been recognized for doing a fantastic job, many admitted that the level of socialization (the primary purpose of the events) significantly suffered. Thus, we conclude that as WFH is not for everyone, WFH mode also falls short of fulfilling all of the software engineers, teams, and organizations' goals. Finally, even though we found WFH to help those recovering from a burnout, we also found that without proper adjustments, remote work can lead to increased exhaustion (long hours of uninterrupted work, fewer breaks, more exhausting virtual meetings), which might have a potential to result in a burnout.

**Table 1: Overview of the WFH practices, ceremonies and routines: changes and novelties**

| Practices and routines | | Adjustments for WFH |
|---|---|---|
| Daily rhythm | Start of work | Many start the day earlier due to a lack of commute time, and/or because they skip the regular morning routines. |
| | Lunch break | Timing changed from a socially imposed "Time for lunch now" to self-imposed "I'll eat when I finish this" and/or "I'll eat when I am hungry". Very few teams hold a joint virtual lunch. For many, the social part of joint lunches is lost (including social conversations by the table and playing video games during lunchtime). |
| | Spontaneous breaks | The number of organic coffee or chit-chat breaks significantly decreased due to the absence of company. |
| | End of work | Changed from leaving the office in regular hours to a "softer" stop and possibly extended work hours, especially in the beginning of the WFH. Some introduce a hard stop as a consequence of working too much. |
| Meetings | One-on-ones | Less individual career development more personal well-being ("How are the things?") |
| | Stand-ups | Extended duration and inclusion of personal status updates and discussions around the member's well-being (typically from 15 to 30 minutes). |
| | Meeting duration | Shifted from ending the meeting earlier to starting a meeting with a short break ("Let's have 5 min to get a coffee) to replace a natural break when walking between the meeting rooms in the office. More meetings run shorter and end on time because of increased fatigue of virtual meetings. |
| | Walk-and-talks | Meetings that do not require computers are increasingly moving from face-to-face or virtual meetings by a computer screen to a phone mediated "walk-and-talk" meetings. |
| | COVID-19 meetings | Meetings clarifying the company policy regarding the COVID-19 and WFH have been held during the first months. Later changed from frequent company-wide meetings to management-only meetings and company-wide emails. |
| | Remote meetings | From divided between the meeting rooms and localized discussions to fully virtual having increased visibility for everybody, i.e., a "level playing field". |
| Work sessions and tactics | Solo work | Became longer with fewer interruptions and increased focus. |
| | Pairing work sessions | Less practiced. Changed from sitting by one computer to virtual sessions with screen sharing and Mural or having parallel work sessions with frequent synchronisation. |
| | Approach to problem solving | Problem solving changed from shouting out "I have a problem" to individual search for solutions or primarily peer-to-peer inquiries via direct messaging. "Always ask team members first" for some drifted to directly contacting the source of knowledge. Increased use of online material over the help from colleagues. |
| | Virtual joint work sessions | A repeating joint virtual work session for the team or a part of the team. Typically, 1-2 hours after the Stand-up, where everyone stays in the virtual meeting room. |
| | Spontaneous discussions | Changed from organic to scheduled discussions. Some engineers schedule meetings promptly (from "let's talk now" to "in 1-2 hours"). Spontaneous discussions facilitated by virtual sessions or drop-in video rooms available all day long. |
| Social activities | Afterworks and social activities | Changed from being regular, well-attended to less frequent and less attended. Afterworks and social activities such as video-gaming, having a coffee or a beer, now changed to the virtual setup. |

## 4.3 Transition From WFH to … WFX

The futuristic future portrayed in the study of WFH pioneers published in 1984 with extensive practice of work from home [1], which was also forecasted in a more recent study of telework [7], seems to be one likely scenario for the post-pandemic times. Based on the attitudes and future expectations of the InterSoft employees who participated in our study, it is evident that many will likely work from home at least a few days a week, and some will move to a new geographic location but stay working in the company. The latter cases have already appeared in InterSoft while writing this article, as the company announced their Work-from-Anywhere (WFX) policy. What can we expect in the future and how WFX will differ from WFH under the COVID-19 times? In contrast to WFH, in which all employees have been equally distributed (which was appreciated), work models in the WFX reality will be numerous. Although employees will be able to make their choices voluntary, "office-home mixes" are likely to strengthen the challenges identified in our study (less pairing, fewer socialization activities, less attended events, more isolated tasks). Besides, the information in partially distributed teams, is likely to circulate in the office clusters and not reach those who work from home [16]. Further, as in sport teams, success is judged both by a player's personal performance and the success of the entire team. An engineer who optimizes only for personal productivity may hurt the productivity of the team. This is why we believe that companies with harmonic team climate and willingness to align the individual decisions to work from anywhere with the interests and needs of a team and perhaps larger parts of the company, are likely to succeed. Similarly, we are concerned with the ability of the companies to sustain their networking and cooperation cultures cultivated through close teamwork and collaboration opportunities. We found that these opportunities become less frequent, rich, and attended when run remotely. Therefore, another important question for the future is identifying the must-happen in-the-office or in-collocation practices, ceremonies, and events that will help maintain the organizational culture.